\documentclass [12pt] {report}
\usepackage{amssymb}
\newcommand {\eqdef} {\stackrel{\rm def}{=}}
\newcommand {\D}[2] {\displaystyle\frac{\partial{#1}}{\partial{#2}}}

\newcommand {\ga} {\gamma}
\newcommand {\la} {\lambda}
\newcommand {\La} {\Lambda}

\newcommand {\Si} {\Sigma}

\newcommand {\fr} {\displaystyle\frac}
\newcommand {\wt} {\widetilde}
\newcommand {\be} {\begin{equation}}
\newcommand {\ee} {\end{equation}}
\newcommand {\ba} {\begin{array}}
\newcommand {\ea} {\end{array}}
\newcommand {\bp} {\begin{picture}}
\newcommand {\ep} {\end{picture}}
\newcommand {\bc} {\begin{center}}
\newcommand {\ec} {\end{center}}
\newcommand {\bt} {\begin{tabular}}
\newcommand {\et} {\end{tabular}}
\newcommand {\lf} {\left}
\newcommand {\rg} {\right}

\newcommand {\cI} {{\cal I}}
\newcommand {\cM} {{\cal M}}

\newcommand {\ses} {\medskip}

\newcommand {\bP} {{\bf P}}

\newcommand {\g}  {\stackrel{g\to -g}{\Longleftrightarrow}}

\newcommand{\mP}{\mbox{$|{\bf P}|$}}

\newcommand {\cP} {{\cal P}}
\newcommand {\bibit} {\bibitem}
\newcommand {\nin} {\noindent}

\def\2#1#2#3{{#1}_{#2}\hspace{0pt}^{#3}}
\def\3#1#2#3#4{{#1}_{#2}\hspace{0pt}^{#3}\hspace{0pt}_{#4}}

\begin {document}

\begin {titlepage}

\vspace{0.1in}

\begin{center}
{\Large Can  Neutrinos and  High-Energy Particles Test Finsler Metric }\\
\end{center}

\begin{center}
{\Large of Space-Time ?}\\
\end{center}

\vspace{0.3in}

\begin{center}

\vspace{.15in} {\large G.S. Asanov\\} \vspace{.25in} {\it Division
of Theoretical Physics, Moscow State University\\ 117234 Moscow,
Russia\\} \vspace{.05in}

\end{center}

\begin{abstract}

The Finsler-relativistic metric function $F(g;R)$  and the
associated Hamiltonian function $H(g;P)$, being considered
together with explicit  Finslerian special-relativistic kinematic
transformations, give rise to a self-consistent  and rigorous
framework upon which corrections to Lorentz-relativistic
quantities can properly be evaluated. The concomitant relations
generalize their Lorentzian prototypes through the presence of a
single characteristic  parameter, ~$g$, so that the explicated
Particle-Antiparticle Asymmetry, as well as the search for
possible distinction between the pseudo-Euclidean Light Geometry
and the Finsler-relativistic Neutrino Geometry, can well be
traced in terms of this parameter. At any fixed rest-mass value
$m$ and $g\ne0$,  the dependences of energy on three-dimensional
momentum prove to be of different forms, as given  by the
respective functions $E^{(+)}(g;m;|{\bf P}|)$ and
$E^{(-)}(g;m;|{\bf P}|)$, for particles and antiparticles. This
splitting of the mass-shell, as well as various  entailed
Finslerian approximations with respect to $g$, can naturally be
proposed for experimental study in order to obtain  estimations on
$g$. Since neutrinos and antineutrinos are uncomposed neat
particles, measuring the difference between their velocities seems
to be the best way for testing implied Finslerian corrections to
conventional Lorentzian quantities. Accelerators which can produce
neutrinos and antineutrinos are ideal instruments to gain this
aim. The known measurements led to the conclusion that the
difference $<0.7\cdot 10^{-4}~(95\%~CL)$, while announced future
long baseline neutrino experiments probably raise the sensitivity
to approach the high level of $\sim 10^{-9}$.

\end{abstract}

\end{titlepage}

\vskip 1cm

{\nin\bf 1. Introduction}
\medskip

{\noindent\bf 1.1. Historical notes}
in front of any relativistic approach should trace back to
the classical Einstein work of 1905 [1] (see also [2--3]), in which
the Special Theory of Relativity (STR) was founded
 upon use of the postulate of light velocity invariance, and to the
great event of 1908 when
 Minkowski reported the adequate relativistic four-dimensional space-time
 geometry based on the pseudo-Euclidean metric function $\sqrt{T^2-X^2}$
 (see [4--5]). The next step of fundamental theoretical significance for STR
was
 made in the work by Frank and Rothe~[6], where the authors argued that the
 assumption of the existence of invariant velocity is not necessary in order
 to arrive at the correct transformations. Since then, various analytical and
 conceptual aspects of the Lorentz-invariant STR have been analyzed in
 numerous keen works (see [7--60] and references therein; many updated
 references can be found in the recent conference volume [61]).

In the domain of particles, the early notes on possible violation of the
STR were made by L.B. Redei, when considering the muon lifetime [62-64].
The ``Redei behaviour" of spectra was taken into account in an extended work
by H.B.Nielsen and coauthors [65-67] devoted to non-Lorentzian aspects.
In this connection, the work by
J.Ellis, M.K.Gailard, D.V.Nanopoulis and S.Rudaz [68] should
also be considered.
The recent work by S.Coleman and S.L.Glashow [69] (continued in [70])
concerned with possible neutrino
tests of the STR would attract new attention of researchers
to possible deeper origin of relativistic invariance.

However, the investigators did
 not invoke the Finsler geometry tools to extend the
 pseudo-Euclidean relativistic framework, so that the metric function
 $\sqrt{T^2-X^2}$ has appeared to be {\it ``a~relativistic invariant of nine
 decades elapsed since 1908 year"}.
\medskip

{\noindent\bf 1.2. Nowadays reasoning} should note, first of all, that
each researcher who makes an attempt to lift the habitual laws and concepts
of the present-day relativistic physics to
the advanced-Finslerian level faces the necessity to explain
in what way he intends to generalize
the ordinary Lorentz-invariance. On the other hand, the
Lorentz-transformations are actually predetermined by the
pseudo-Riemannian patterns adopted to metricize the space-time
base manifold. Therefore, the very possibility of the
Euclidean-to-Finslerian extension of the Lorentz-invariant
relativistic theories depends crucially on whether the
Finsler-relativistic geometry for the space-time
can be constructed in a consistent
and trustworthy way. It proves that the latter-type
geometry can well be founded upon the use of the
Finslerian metric function (FMF) $F(g;R)$ presented below and of the associated
Finslerian metric tensor (FMT) $g_{pq}(g;R)$ [71--88].
\medskip

{\noindent\bf 1.3. The uniqueness theorem for the special-relativistic FMF}
comes from a due attentive consideration.
Indeed, let $M$ be the background four-dimensional space-time, $R\equiv
\{R^p\}\in M; p,q,\dots=0,1,2,3$. Any FMF
 $F (R)$ defines the hypersurface $\cI=\{R\in M:F(R)=1\}$ called
conventionally \it the indicatrix  \rm (see [89--94]). The well-known fact is that the
 pseudo-Euclidean geometry may be characterized by the condition that the
 associated indicatrix is a pseudosphere of radius~1 (a hyperboloid),
and hence is a space of
constant curvature~$-1$. The case founded on the property
\bigskip

(P1) {\it The indicatrix is a space of constant negative curvature}
 $R_{\cI}\ne-1$\\[2mm]
 can be regarded as the nearest Euclidean-to-Finslerian relativistic
 generalization. Also, we assume the properties\ses

 (P2) {\it The FMF is compatible with the principle of spatial isotropy
(the $\cP$-parity),}

 (P3) {\it The associated FMT is of the time-space signature $(+---)$,}\ses

\noindent and\ses

 (P4) {\it The principle of correspondence holds true,}\\[1mm]
 that is, the associated FMT reduces exactly to its ordinary known
special-relativistic prototypes when $R_{\cI}\to-1$, which physical
 significance is quite transparent.
\ses

 All the items (P1)--(P4) are obeyed whenever the choice $F= F(g;R)$
 is made, in which case
$$
 R_{\cI}=-\lf(1+\fr14g^2\rg).
$$
 Vice versa, we can claim the following\ses

THEOREM \it The properties
\rm(P1)--(P4), \it when treated as conditions imposed on
the FMF, specify it unambiguously in the form $F=F(g;R)$.
\rm
\medskip
For more deatil, the reader is referred to [87].
\medskip

{\noindent\bf 1.4. The  special-relativistic FMF}
$$
F(g;R)=\lf|T+g_-|{\bf R}|\rg|^{G_+/2}\lf|T+g_+|{\bf R}|\rg|^{-G_-/2}
$$
can be adduced by the Hamiltonian function
$$
H(g;P)=\lf|P_0-\fr{|{\bP}|}{g^+}\rg|^{G^+/2}
\lf|P_0-\fr{|{\bP}|}{g^-}\rg|^{-G^-/2},
$$
where the following notation has been used:
$$
\qquad h\eqdef\sqrt{1+\fr14g^2},
$$
$$
g_+=-\fr12g+h, \qquad g_-=-\fr12g-h,
$$
\medskip
$$
g^+=1/g_+=-g_-,  \qquad  g^-=1/g_-=-g_+,
$$
\medskip
$$
g^+=\fr12g+h, \qquad g^-=\fr12g-h.
$$
\medskip
$$
G_+=g_+/h, \quad G_-=g_-/h,
$$
\medskip
$$
G^+=g^+/h, \quad G^-=g^-/h,
$$
together with $R^0=T$. Notice that
$$
g_+\g -g_-, \qquad g^+\g -g^-, \qquad G_+\g -G_-, \qquad G^+\g -G^-.
$$
Various relativistic consequence of use of these $F(g;R)$ and $H(g;P)$
have been studied in [71--88].
\medskip

{\noindent\bf 1.5. The characteristic Finslerian parameter \boldmath$g$} comes
 directly from the indicatrix curvature
$R_{\cI}$. The question as to what is the
 physical sense and meaning of the parameter~$g$ can be answered by focusing
 attention on varios lucid Finsier-relativistic relations,
 thereby giving rise to quite a number of
 fairly unexpected juxtapositions.
For instance, noting the formulae given below in Section 2,
we can conclude that there are no Finslerian corrections to
 the velocity-momentum transition if and only if there are no Finslerian
 corrections to the law of addition of relativistic velocities.

 Characteristically, the difference of the indicatrix curvature from $-1$
 shows itself beginning with the second order of the parameter~$g$,
whereas the FMT components $g_{pq}(g;R)$ start differring
from the pseudo-Euclidean diag$\{1,-1,-1,-1\}$ in the first order.

 The degree of smallness of the input parameter $g$ might be estimated
 by using experimental tests of possible violation of the traditional
 Lorentz invariance. Various relations
can be proposed to use in such tests.
The important and urgent questions are arising:
\medskip

\it Is today's experimentaly accuracy sufficient to predict the parameter~$g$?

 How to disantangle the parameter~$g$ from extended relativistic experience?

\rm\noindent
The value of $g$ should characterize the degree to which
Lorentz invariance is broken in nature.
\medskip

{\noindent\bf 1.6. ``Universal'' means ``geometrical''.} The Finslerian parameter $g$
 is not ``bound up" to any particular type of fundamental physical interactions,
 although the corrections in $g$ may enter the equations describing any of
 the fundamental interactions (through, for example, $g$-corrections to the
 metric tensor, tetrads, and connection coefficients). The Finslerian
 approach does not assume that some fundamental length should be introduced,
 although there exist numerous analogies among the Finslerian consequences
 and the consequences of the elementary length-based theories (for example,
 the occurrence of non-Lorentzian corrections). The parameter~$g$ is not a
 combination of well-known fundamental physical constants, but, instead, is
 meant to be a dimensonless fundamental constant. The parameter~$g$ is
 universal in the sense that it is of pure geometrical origin, that is, the
 corrections to the pseudo-Riemannian geometry of space-time are introduced
 through this parameter~$g$. Merely, \it the parameter~$g$ evaluates the degree
 of Finslerian non-Riemannianity of space-time.\rm
\medskip

{\noindent\bf 1.7. Non-Lorentzian transformations}  were
considered in several works.
The Lorentz transformations and their modifications have been
 serving over our century to ``work-up" the high-energy phenomenology,
 derive the fundamental physical field equations, and
 predict new relativistic effects. Despite the general feeling of a high
 degree of accuracy between predictions and measurements, various
 modifications, including the well-known cases\ses

 (I) The Robertson Transfomations [10]

 (II) The Edward Transformations [18, 22]

 (III) The Mansouri--Sexl Transformations [30]

 (IV) The Tangherlini Transfornations [16]\ses

\noindent
 (listed here in the chronological order) have been used \big[the
 transformations (I)--(III) have clearly been compared with each other in~[39];
 a~systematic review of various kinematics relations stemed from the choice
 (IV) can be found in~[28]\big]. Surprisingly, in a sharp contrast to the
 approach followed in Einstein's work~[1,~2], in the which STR began with two
 fundamental invariance principles to derive the required transformations, a
 lack of profound invariance motivation to favour relativistic treatments is
 a common feature for the approaches based on the non-Lorentzian
 transformations (I)--(IV). In fact, the transformations (I)--(IV) have been
 introduced primarilly to reanalyze the role of synchronization procedure
 [11, 14,~15,~60--61]. No metric function invariant
 under a member of the set of non-Lorentzian transformations (I)--(IV) has
 been known.
\bigskip

{\nin\bf 2. The Finslerian Extension of Lorentz Transformations}
\bigskip

 However, the geometrically-motivated invariance can be retained
 safely if the Finsler geometry is invoked as a necessary basis thereto.
Such a way leads to the following Finslerian special-relativistic
kinematic transformations
$$
R^0=\La_0^0\wt R^0+\La_1^0\wt R^1, \quad R^1=\La_0^1\wt R^0+\La_1^1\wt R^1,
$$$$
R^2=\La_2^2\wt R^2, \quad R^3=\La_3^3\wt R^3,
$$
where the coefficients are
$$
\La_0^0=1/V(g;v), \quad \La_0^1=\La_1^0=v/V(g;v), \quad \La_1^1=(1-g|v|)/V(g;v),
$$
$$
\La_2^2=\La_3^3=\sqrt{Q(g;v)}/V(g;v).
$$
The transformations can be inversed to give
$$
\wt R^0=\la_0^0R^0+\la_1^0R^1, \quad \wt R^1=\la_0^1R^0+\la_1^1R^1,
$$
$$
\wt R^2=\la_2^2R^2, \quad \wt R^3=\la_3^3 R^3
$$
with
$$
\la_0^0=(1-g|v|)V(g;v)/Q(g;v), \quad
\la_0^1=\la_1^0=-vV(g;v)/Q(g;v),
$$
$$
\la_1^1=V(g;v)/Q(g;v),
$$
$$
\la_2^2=\la_3^3=V(g;v)/\sqrt{Q(g;v)}.
$$
The co-version transformations, as related to a four-dimensional momentum
$P_p=\{P_0,P_a\}$, read
$$
P_0={\La^*}^0_0\wt P_0+{\La^*}^1_0\wt P_1, \quad
P_1={\La^*}_1^0\wt P_0+{\La^*}_1^1\wt P_1,
$$$$
P_2={\La^*}_2^2\wt P_2, \quad P_3={\La^*}_3^3\wt P_3
$$
with
$$
{\La^*}_0^0=1/W(g;p), \quad {\La^*}_1^0={\La^*}_0^1=-p/W(g;p),
$$
$$
{\La^*}_1^1=(1+g|p|)/W(g;p),
$$
\medskip
$$
{\La^*}_2^2={\La^*}_3^3=\sqrt{Q^*(g;p)}/W(g;p),
$$
and for their inverse,
$$
\wt P_0={\la^*}_0^0P_0+{\la^*}_0^1P_1,
\quad \wt P_1={\la^*}_1^0P_0+{\la^*}_1^1P_1,
$$$$
\wt P_2={\la^*}_2^2P_2, \quad \wt P_3={\la^*}_3^3P_3,
$$
where
$$
{\la^*}_0^0=(1+g|p|)W(g;p)/Q^*(g;p),
\quad {\la^*}_1^0={\la^*}_0^1=pW(g;p)/Q^*(g;p),
$$
$$
{\la^*}_1^1=W(g;p)/Q^*(g;p),
$$$$
{\la^*}_2^2={\la^*}_3^3=W(g;p)/\sqrt{Q^*(g;p)}.
$$
Here, $V=F/R^0$,  $W=H/P_0$, and
$$
Q= 1-g|v|-v^2,\qquad Q^*= 1+g|p|-p^2;
$$
the notation $v$ and $p$ has been used for the three-dimensional
velocity and momenta, respectively.
Invariance
$$
F(g;R) = F(g;\wt R), \qquad H(g;P) = H(g;\wt P)
$$
holds true.
{\it In the limit $g \to 0$ the above transformations turn into
the conventional Lorentz transformations}.

It should be noted that
$$
\La_2^2\ne1 \quad \& \quad \La_3^3\ne1 \quad whenever \quad g\ne0
$$
so that we cannot get any generalization if,
attempting to generalize the Lorentz transformations in the
Finslerian way, we try to retain ``as obvious" the principle stating that
``the scales perpendicular to motion direction must not be deformed".

The above formulae  entail, in particular, that {\bf the Finslerian
addition law for relative velocities} should read
$$
v_3=v_2\oplus v_1:\qquad v_3=\fr{v_2+v_1-gv_2|v_1|}{1+v_2v_1}.
$$
The inverse subtraction law is
$$
v_2=v_3\ominus v_1:\qquad v_2=\fr{v_3-v_1}{1-v_1v_3-g|v_1|}
$$
which entails the relations
$$
\fr1{v_3\ominus v_1}+\fr1{v_1\ominus v_3}=
g\fr{|v_3|-|v_1|}{v_3-v_1}
$$
and
$$
\ominus v=-\fr v{1-g|v|}
$$
together with the \it fundamental group property \rm
$$
(v_1\oplus v_2)\oplus v_3=v_1\oplus(v_2\oplus v_3).
$$
Quite similar Finslerian observations can be made for relativistic momenta.

We observe that the Finslerian extension  violates the reciprocity principle
(which claims that the velocity of a inertial reference frame
$S$ measured from another inertial reference frame $S'$ is
the opposite of the velocity of $S'$ measured from~$S$), for
$$
\ominus v\ne-v\quad \rm whenever \it \quad g\ne0.
$$
Owing to the remarkable group property,
we are justified in claiming that the Finslerian extension of STR
sprung from the FMF $F(g;R)$ does obey the
requirement that the kinematic transformations linking pairs of
observers must form a group (see more detail related to the
material of the present section in [88]).
\bigskip

{\nin\bf 3. Energy-momentum dependence of respective Finsler type}
\medskip

Considering the Finsler-mass-shell
$$
H(g;P_0,\mP)=m,
$$
where $m$ is the rest mass of the particle,
separately in the future-like sector and in the past-like
sector, we obtain the two sheets,
$\cM^{(+)}(m)$ and $\cM^{(-)}(m)$ ,
defined  respectively by the equations
$$
H^{(+)}(g;P_0,|{\bf P}|)=m, \qquad H^{(-)}(g;P_0,|{\bf P}|)=m,
$$
which give rise to the energy-momentum functions
\bigskip
$$
P_0=P_0^{(+)}(g;m;|{\bf P}|)>0,\qquad
P_0=P_0^{(-)}(g;m;|{\bf P}|)<0.
$$
It occurs that, as far as we venture to follow the
Finsler-relativistic approach, the sheet $\cM^{(-)}(m)$
{\it  ceases to be the mirror image of the sheet} $\cM^{(+)}(m)$

Considering the energies
$$
E^{(+)}=
P_0^{(+)}>0,\qquad E^{(-)}=-P_0^{(-)}>0,
$$
we can find after direct calculations the simple result:
$$
\D{E^{(+)}}{|{\bf P}|}=\fr{|{\bf P}|}{E^{(+)}+g|{\bf P}|},
\qquad
\D{E^{(-)}}{|{\bf P}|}=\fr{|{\bf P}|}{E^{(-)}-g|{\bf P}|},
$$
These formulae can be used to conclude that, nevertheless,
$E^{(+)}$ and $E^{(-)}$ are {\it monotonically increasing functions of}
$|{\bf P}|$.
The symmetry
$$
E^{(+)}\g E^{(-)}
$$
holds now, instead of the ordinary Lorentzian identity
$E^{(+)}= E^{(-)}$.

Whenever $m>0$, we can put
$|{\bf k}|=\mP/m$ and examine the low-velocity approximation
$$
|{\bf k}|\ll m,
$$
which yields the following differing energy-momentum dependences for the case
of a particle and for the case of an antiparticle:
$$
E^{(+)}=1+\fr12|{\bf k}|^2-\fr13g|{\bf k}|^3-\fr1{24}(3+4g^2)|{\bf k}|^4+{\rm O}(g|{\bf k}|^5)
$$
and
$$
E^{(-)}=
1+\fr12|{\bf k}|^2
+\fr13g|{\bf k}|^3-\fr1{24}(3+4g^2)|{\bf k}|^4+{\rm O}(g|{\bf k}|^5)
$$
which can be subjected to a due experimental verification,
at least in principle, to get estimations on the parameter $g$.
As a neat particular consequences, we obtain
$$
E^{(+)}-E^{(-)}=-\fr23g|{\bf k}|^3,
$$
to the nearest Finslerian order.
\bigskip

From this it follows that we should deal with the two Finsler-relativistic
Hamiltonian functions
$$
H_1(g;E^{(+)},|{\bf P}|)=
\Bigl|E^{(+)}-\fr{|{\bf P}|}{g^+}\Bigr|^{G^+/2}
\Bigl|E^{(+)}-\fr{|{\bf P}|}{g^-}\Bigr|^{-G^-/2}
$$
and
$$
H_2(g;E^{(-)},|{\bf P}|)=
\Bigl|E^{(-)}+\fr{|{\bf P}|}{g^+}\Bigr|^{G^+/2}
\Bigl|E^{(-)}+\fr{|{\bf P}|}{g^-}\Bigr|^{-G^-/2},
$$
the respective
relevant
energy-momentum relations
being governed by the equations
$$
H_1(g;E^{(+)},|{\bf P}|)=m
, \qquad
H_2(g;E^{(-)},|{\bf P}|)=m.
$$
Thus we observe the phenomen of  {\it the Finslerian splitting} of
the ordinary Lorentzian mass-shell, for
\medskip

\nin \it the mass-shells defined in terms of $H_1$, and in terms of $H_2$,
differ from one another unless $g=0$.\rm
\medskip

The fact is that the Finsler-relativistic Hamiltonian function $H(g;P)$
written out in Section 1.4 is no more $P_0-$ even:
$$
H(g;-P_0,\bP)\ne H(g;P_0,\bP),
\qquad unless \quad g=0.
$$
Instead, the function  shows the property of
$gT$-\it parity\rm
$$
H(-g;-P_0,\bP)= H(g;P_0,\bP),
$$
in addition to the property of $\cP$-\it parity\rm
$$
H(g;-P_0,-\bP)= H(g;P_0,\bP)
$$
that is retained under the given Finslerian extension.

Thus the relativistic
Finslerian approach
substitutes {\it the combined}
$gT${\it -symmetry} with the ordinary, Lorentzian and  pseudo-Euclidean,
$T${\it-parity}.

It can be said also that the Finslerian parameter $g$ measures
the degree of the respective Particle-Antiparticle asymmetry.
\bigskip

{\nin\bf 4. New Call to Experimenters: Finslerian Neutrino Geometry
\it\bf Versus \rm\bf
Lorentzian Light Geometry ?}
\medskip

At present it has become customary to treat the mean rest frame $\Si$
of the universe (the microwave background frame) as the
preferred
reference frame and use the Earth's speed $\sim300-400$~km$\cdot$s$^{-1}$
with respect to~$\Si$
in evaluating estimations for phenomenological parameters of
possible violation of the relativistic quantities from
their traditional Lorentzian patterns (relevant experimental tests,
which are often based on the usage of the modern high-precision laser
techniques, are many (see [41--57]). Properly, S.Coleman and S.L.Glashow
in the known work [69], have begun with
noting a preferred frame
to study possible ``Neutrino Tests of Special Relativity".

Moreover, many authors prefer to identify the
frame $\Si$ with the Cosmic Substratum, noting that our phenomenological
situation in Cosmos just compels attention to a distinct possibility
of such media (see [61]).
Also, such a media is seemingly an ideal ground for the notion of
Cosmic Vacuum Media formed at any local point by averaging over all physical
quantum or stochastic fields (similar ideas can be ``projected down" from
the modern multi-dimensional string theories; see in this respect V.Ammosov and
G.Volkov [97] and references therein).

On the other hand, any extension of pseudo-Euclidean square-root metric
should obviously affect and deform
the Lorentzian mass-shell for high-energy particles.
Whence the relativistic and ultra-relativistic
particles should be sensitive to the primary geometry
of space-time.
Especially, since the mass-shell deformation is accompanied by due
deformation of the light cone, we should expect that
the neutrinos and antineutrinos, as being entirely uncomposed
 particles,
do feel any possible Finsler-correction \it in an utmost and neat way\rm.
\it But the light photons don't at all\rm.

It can be predicted, therefore,
that the neutrinos are quite sensitive to such corrections.
Granted the metric is Finslerian rather than pseudo-Euclidean,
the neutrinos live in a Finsler geometry rooted in Space-Time
\it versus \rm the light photons which, because  there are no anti-photons,
follows  monotonically the pseudo-Riemannian
future-oriented cosmic substratum in a sense.

As we pointed out in the preceding section, an accurate-way
Finslerian approach entails eventually the asymmetry between
the up- and down-sheets of the mass-shell such that, at any fixed rest-mass,
\medskip

\nin\it the mass-shell of a particle is not identical to the mass-shell
of its antiparticle
unless the Finslerian characteristic parameter, $g$, is null\rm.
\medskip

Thus the Finsler-relativistic mass-shell at any fixed rest-mass value is now
splitting into two different sheets, as being related to particles and
antiparticles,
so that, on an accurate Finsler-relativistic extension, the Neutrinos and
the Antineutrinos occupy some comfortable twin-compartments in a
Finsler-relativistic train, while the Photon travels are left intact and
outside. \it There is no single compartment in the train.\rm
\bigskip

{\nin\bf 5. Direct Experimental Claim: the neutrino velocity may differ
from the light velocity?}
\medskip

\it To the first order of magnitude
with respect to the Finslerian parameter \rm $g$,
$$
|g|<<1,
$$
the constants written down in Section 1.4 reduce to
$$
G_+=g_+=1-\fr g2, \quad G_-=g_-=-1-\fr g2,
$$
$$
G^+=1+\fr g2, \quad G^-=-1+\fr g2,
$$
so that the squared Finsler-relativistic Hamiltonian function takes on
the form
$$
[H(g;P)]^2=\lf|P_0-(1-\fr g2){|{\bP}|}\rg|^{1+\fr g2}
\lf|P_0+(1+\fr g2){|{\bP}|}\rg|^{1-\fr g2}
$$
\bigskip\\
that is broken into two cases
$$
[H_1(g;P)]^2=\lf(E-(1-\fr g2){|{\bP}|}\rg)^{1+\fr g2}
\lf(E+(1+\fr g2){|{\bP}|}\rg)^{1-\fr g2}
$$
and
$$
[H_2(g;P)]^2=\lf(E+(1-\fr g2){|{\bP}|}\rg)^{1+\fr g2}
\lf(E-(1+\fr g2){|{\bP}|}\rg)^{1-\fr g2},
$$
extending the ordinary ansatz
$$
[H_{pseudo-Euclidean}]^2=\lf(P_0-{|{\bP}|}\rg)\lf(P_0+{|{\bP}|}\rg)
$$
in an accurate Finslerian $O(g)$-way.

If the function $H_1$ relates conventionally to particles, then $H_2$
should relate to antiparticles.

Thus, the account for the
Finslerian parameter $g$ does
shift the ordinary isotropic cone $\Si:\,P_0=\pm\mP$,
leading actually to the unification
$\Si^+\cup\Si^-$ which members, $\Si^+$ and $\Si^-$,
don't mirror one another, for
$$
\Si^+:\quad P_0=(1-\fr g2)\mP
$$
and
$$
\Si^-:\quad P_0=-(1+\fr g2)\mP.
$$

\it How should this splitting  display in the elementary particle physics? \rm

Following the ordinary relativistic treatment of the antiparticle
hyperboloid to be the reflection of the past-like
hyperboloid into the future region, we ought to recognize
that in the Finslerian framework the rest-mass antiparticles should
correspond to the cone
$$
(\Si^+)^*\eqdef-\Si^-:\quad
E=(1+\fr g2)\mP,
$$
which differs from the rest-mass particle-kind cone proper
$$
\Si^+:\quad E=(1-\fr g2)\mP.
$$

This entails immediately the important conclusion
that, on calibrating away the
speed-of-light to be unity,
the velocity values
$$
v\eqdef\mP/E
$$
for neutrinos and antineutrinos should
differ from one another according to the rule:
$$
v_{\nu}=1+\fr g2, \quad v_{\bar\nu}=1-\fr g2.
$$

The middle sum remains intact
$$
\fr{v_{\bar\nu}+v_{\nu}}2=1,
$$
but the velocity difference
$$
v_{\nu}-v_{\bar\nu}=g_{\nu}.
$$
is significant for careful experimental verifications.

In this respect, it will be noted that
the recent particle accelerator data seem to support everywhere
the estimation
$g_{\nu}<10^{-4}-10^{-5}$ for all charged long-lived elementary particles.
The velocity
for accelerator-produced
neutrinos
was studied for the first time in the  experiments  performed
at the  Fermilab [95--96], which resulted in
the  following  upper  limit at  95\%  CL:
$$
|v_{\nu}/c - v_{\bar\nu}/c| < 0.7\cdot 10^{-4}.
$$
As was pointed out recently in the work by V.Ammosov and G.Volkov [97] ,
the expected pending neutrino experiments can get  sensitive enough to
measure the neutrino-antineutrino velocity differencies
 $\sim 10^{-6}$ for short baseline experiments
and $\sim 10^{-9}$ for long baseline experiments,
and that these sensitivities can directly
transform into the $g_{\nu}$ sensitivity. Since, as we have argued in
Section 4, the photons $\ga$ don't feel any Finsler geometry, we have
for them formally $g_{\ga}=0$.

Certainly, the STR cannot consent to draw any distinction between the
neutrino velocity and the light velocity (in vacuum). Whence any
positive outcome of such-type experiments would put reliable limits
on all the body of the STR.

\def\bibit[#1]#2\par{\rm\noindent\parskip1pt
                     \parbox[t]{.05\textwidth}{\mbox{}\hfill[#1]}\hfill
                     \parbox[t]{.925\textwidth}{\baselineskip11pt#2}\par}

\noindent{\bf References}
\bigskip

\bibit[1] A. Einstein: \it Ann. Physik \bf17 \rm(1905), 891.

\bibit[2] A. Einstein: \it The Meaning of Relativity, \rm 5th ed., Princeton, 1955.

\bibit[3] W. Perret and G.B. Jeffery: \it The Principle of Relativity,
\rm Dover, N.\,Y., 1958.

\bibit[4] H. Minkowski: \it Raum und Zeit~--- Phys. Z. \bf10 \rm(1909), 104.

\bibit[5] H. Minkowski: \it Das Relativit\"atsprinzip~--- Ann. Physik. \bf47
\rm(1915), 927.

\bibit[6] Ph. Frank and H. Rothe: \it Ann. Physik \bf34 \rm(1911), 825.

\bibit[7] A.S. Eddington: \it The Mathematical Theory of Relativity,
\rm Cambridge University Press, Cambridge, 1924.

\bibit[8] E.A. Milne: \it Kinematic Relativity, \rm Oxford University Press,
Oxford, 1948.

\bibit[9] J.L. Synge: \it Relativity: The Special Theory, \rm
 North-Holland, Amsterdam 1956.

\bibit[10] H.P. Robertson: \it Rev. Mod. Phys. \bf 21 \rm(1949), 378.

\bibit[11] H. Reichenbach: \it The Philosophy of space and time, \rm Dover Pupl., Inc.,
N.\,Y., 1958.

\bibit[12] B. Jaff\'e: \it Michelson and the Speed of Light, \rm
Anchor Books, Doubleday, N.\,Y., 1960.

\bibit[13] H. Bondi: \it Rept. Progr. Phys. \bf22 \rm(1959), 97.

\bibit[14] A. Gr\"unbaum: \it The Philosophy of Science, \rm A.~Danto and S.~Morgenbesser,
eds., Meridian Books, N.\,Y., 1960.

\bibit[15] A. Gr\"unbaum: \it The Philosophy of Space and Time,
\rm Redei, Dordrecht, 1973.

\bibit[16] F.R. Tangherlini: \it Suppl. Nuovo Cimento \bf20 \rm(1961), 1.

\bibit[17] M. Ruderfer: \it Phys. Rev. Lett. \bf 5 \rm(1960), 191;
\it Proc. IRE, \bf48, \rm(1960), 1661; \bf50 \rm(1962), 325.

\bibit[18] W.F. Edwards: \it Am. J. Phys. \bf31 \rm(1963), 482.

\bibit[19] Ph. Tourrenc, T. Melliti, and J. Bosredon: \it GRG \bf28 \rm(1996),
1071.

\bibit[20] V. Berzi and V. Gorini: \it J. Math. Phys. \bf10 \rm(1969), 1518.

\bibit[21] H.M. Schwartz: \it American J. Phys. \bf30 \rm(1962), 697;
\bf39 \rm(1971), 1283; \bf40 \rm(1972), 862; \bf52 \rm(1984),~346.

\bibit[22] J.A. Winnie: \it Philos. Sci. \bf37 \rm(1970), 81, 223.

\bibit[23] A. Ungar: \it Philos. Sci. \bf 53 \rm(1986), 395.

\bibit[24] A.P. French: \it Special Relativity, \rm MIT Press,
 Norton, N.Y., 1968.

\bibit[25] R. Torretti: \it Relativity and Geometry, \rm Pergamon Press, 1983.

\bibit[26] F. Goy: \it Found. Phys. Lett. \bf 9\rm(2) (1996), 165.

\bibit[27] E.A. Desloge: \it Found. Phys. \bf 19 \rm(1989), 1191.

\bibit[28] G. Spavieri: \it Found. Phys. Lett. \bf 1 \rm (1988), 373.

\bibit[29] S.J. Prokhovnik and W.T. Morris: \it Found. Phys.
\bf 19 \rm(1989), 531.

\bibit[30] R. Mansouri and R. Sexl: \it Gen. Rel. Grav. \bf8 \rm(1977), 496, 515, 809.

\bibit[31] S.J. Prokhovnik:
\it Found. Phys. \bf 3 \rm(1973), 351; \bf9
\rm(1979), 883; \bf10 \rm(1980, 197; \bf19 \rm(1989), 541.

\bibit[32] S.J. Prokhovnik:
\it J.~Australian
Math. Soc. \bf5\rm(2) (1965), 273; \bf6\rm(1) (1966), 101.

\bibit[33] S.J. Prokhovnik: \it The Logic of Special Relativity, \rm
Cambridge University Press, Cambridge, 1967.

\bibit[34] A.K.A. Maciel and J. Tiomno: \it Found. Phys. \bf 19 \rm(1989), 505 and 521.

\bibit[35] G. Spavieri: \it Phys. Rev. \bf A34 \rm(1986), 1708.

\bibit[36] C.I. Mocanu: \it Found. Phys. Lett. \bf 5 \rm(1992), 443.

\bibit[37] W.A. Rodrigues and J. Tiomno: \it Found. Phys. \bf 15 \rm(1985), 945.

\bibit[38] W.H McCrea: \it Proc. Math. Soc. Univ. Southampton \bf5 \rm(1962), 15.

\bibit[39] Y.Z. Zhang: \it Gen. Rel. Grav. \bf27 \rm(1995), 475.

\bibit[40] F. Selleri: \it Found. Phys. Lett. \bf 9\rm(1) (1997), 73;
\it Found. Phys. \bf26 \rm(1996), 641.

\bibit[41] D.G. Torr and P. Kolen: \it Found. Phys. \bf 12 \rm(1982), 256 and 401.

\bibit[42] T. Chang: \it Phys. Lett. \bf 70A \rm(1979),~1; \it J.~Phys.
\bf A13 \rm(1980), L207.

\bibit[43] M.P. Haugan and C.M. Will: \it Physics Today \bf 40 \rm(1987), 69.

\bibit[44] C.M. Will: \it Phys. Rev. \bf D45 \rm(1992), 403.

\bibit[45] S. Marinov: \it Czech. J. Phys. \bf24 \rm(1974), 965; \it Found. Phys.
\bf9 \rm(1979), 445; \it Gen. Rel. Grav. \bf12 \rm(1980), 57.

\bibit[46] E.M. Kelly: \it Found. Phys. \bf14, \rm(1984), 705; \bf15 \rm(1985), 333.

\bibit[47] A. Brillet and J.L. Hall: \it Phys. Rev. Lett. \bf 42 \rm(1979), 549.

\bibit[48] M. Kaivola et al.: \it Phys. Rev. Lett. \bf 54 \rm(1985), 255.

\bibit[49] T.P. Krisher et al.: \it Phys. Rev. \bf D42 \rm(1990), 731.

\bibit[50] D. Hils and J.L. Hall: \it Phys. Rev. Lett. \bf 64 \rm(1990), 1697.

\bibit[51] J. M\"uller and M.H. Soffel: \it Phys. Lett. \bf 198A \rm(1995), 71.

\bibit[52]
E. Fischbach, M.P. Haugan, D. Tadic, and H.-Y. Cheng: \it
Phys. Rev. \bf D32 \rm(1985), 154.

\bibit[53] G.L.Greene, M.S.Dewey, E.G.Kessler,Jr., and
E. Fischbach: \it
Phys. Rev. \bf D44 \rm(1991), 2216.

\bibit[54] V.W.Hughes, H.G.Robinson, and V.Beltran-Lopez:
\it Phys. Rev. Lett. \bf 4 \rm(1960), 342.

\bibit[55] R.W.P.Drever: \it Philos.Mag. \bf 6 \rm(1961), 683.

\bibit[56] J.D.Prestage, J.J.Bollinger, W.M.Itano, and D.J.Wineland:
\it Phys. Rev. Lett. \bf 54 \rm(1985), 2387.

\bibit[57] S.K.Lamoreaux, J.P.Jacobs, B.R.Heckel, F.J.Raab,
and E.N.Fortson: \it Phys. Rev. Lett. \bf 57 \rm(1986), 3125.

\bibit[58] J. M\"uller and M.H. Soffel: \it Phys. Lett. \bf 198A \rm(1995), 71.

\bibit[59] E.T. Whittaker: \it History of the Theories of Aether and Electricity:
the Classical Theories, \rm Nelson, London, 1962.

\bibit[60] R. Anderson, I.Vetharaniam, and G.E.Stedman: Conventionality of
synchronisation, gauge dependence and test theories of
relativity. \it Phys. Reports \bf 295 \rm(1998), 93-180.

\bibit[61]
Proceedings of Conference \it``Physical Interpretation of
Relativity Theory" \rm, London, Sunderland, 2000.

\bibit[62]
L.B.Redei:
Possible experimental
tests of the existence of a universal length, \it Phys.Rev. \bf 145
\rm(1966), 999.

\bibit[63]
L.B.Redei: Validity of special relativity at small distances and the
velocity dependence of the muon lifetime, \it Phys.Rev. \bf 162 \rm(1967), 1299.

\bibit[64]
I.-E.Lundberg and L.B.Redei: Validity of special relativity at small distances and the
velocity dependence of the charged-pion lifetime, \it Phys.Rev. \bf162
\rm(1967) 1299; \it Phys.Rev. \bf169 \rm(1968) 1012.

\bibit[65]
H.B.Nielsen and I.Picek: Lorentz non-invariance, \it Nuclear Phys. \bf B211
\rm(1983), 269-296.

\bibit[66]
H.B.Nielsen and I.Picek:  Lorentz invariance as a low energy phenomenon,
\it Nuclear Phys. \bf B217 \rm(1983), 125-144.

\bibit[67]
H.B.Nielsen and I.Picek: The Redei-like model and testing
Lorentz invariance. \it Phys. Lett. \bf114B, \rm N 2,3 (1982), 141-146.

\bibit[68] J.Ellis, M.K.Gailard, D.V.Nanopoulis and S.Rudaz: \it Nucl.
Phys. \bf B176 \rm(1980), 61.

\bibit[69] S.Coleman and S.L.Glashow: Cosmic ray and neutrino tests of
special relativity.
\it Harvard University Report \rm No. HUTP-97/A008, hep-ph/9703240.

\bibit[70] S.L.Glashow, A.Halprin, P.L.Krastev, C.N.Leung, and Pantaleone:
Remarks on  neutrino tests of special relativity. \it Phys. Rev. \bf D56
\rm ,N 4 (1997), 2433-2434.

\bibit[71] G.S. Asanov:
Finslerian relativistic space compartible with space isotropy.
 \it Moscow University Physics Bulletin \bf 35 \rm(1) (1994), 19.

\bibit[72] G.S. Asanov:
Finslerian corrections to probability of decay {$T=P+Q$}.
 \it Moscow University Physics Bulletin \bf 35 \rm(5) (1994), 3.

\bibit[73] G.S. Asanov:
Finslerian extension of Lorentz transformations.
 \it Moscow University Physics Bulletin \bf 35 \rm(4) (1995), 7.

\bibit[74] G.S. Asanov:
Finsler cases of GF-spaces, \it Aequationes Math.,
\bf 49 \rm(1995), 234.

\bibit[75] G.S. Asanov:
 Finslerian kinematic consequences.
\it Moscow University Physics Bulletin
\bf 35 \rm (1) (1996), 18.

\bibit[76] G.S. Asanov.
Finslerian transformation laws for light signal velocities.
\it Moscow University Physics Bulletin
\bf 37 \rm(3) (1996), 8.

\bibit[77] G.S. Asanov:
Finslerian extension of scalar products.
\it Moscow University Physics Bulletin
\bf 37 \rm(4) (1996), 3.

\bibit[78] G.S. Asanov:
 Finslerian non-linear invariance and Lorentz transformations.
\it Moscow University Physics Bulletin
\bf 37 \rm(2) (1996), 8.

\bibit[79] G.S. Asanov:
Finslerian metric and tetrads in static spherically-symmetric
  case of gravitational field \it Reports on Math. Phys.\bf 39
  \rm(1997), 69-75.

\bibit[80] G.S. Asanov:
 Finslerian {$g-$}correction of  relativistic dynamic relations.
\it Moscow University Physics Bulletin
\bf 39 \rm(5) (1998), 3.

\bibit[81] G.S. Asanov:
Finslerian invariant and coordinate length in inertial reference frame.
\it Moscow University Physics Bulletin
\bf 39 \rm(1) (1998), 18.

\bibit[82] G.S. Asanov:
Finslerian approach to theory of quantized fields. Positivity of energy
of scalar fields and generalization of Pauli-Jordan functions.
\it Moscow University Physics Bulletin \bf 39 \rm(3) (1998), 15.

\bibit[83] G.S. Asanov:
Finslerian metric functions over the product {$R\times M$} and
  their potential applications.
\it Reports on Math. Phys.\bf 41 \rm(1998), 117-132.

\bibit[84] G.S. Asanov:
Finslerian extension of Lorentz transformations.
\it Reports on Math. Phys.\bf 42 \rm(1998), 273-296.

\bibit[85] G.S. Asanov:
Finsler space {${\cal F}_{PD}$} of positive-definite type
  gives rise to a lenticular extension of {M}axwell's distribution law.
\it Reports on Math. Phys., \bf 43 \rm(199), 437-465.

\bibit[86] G.S. Asanov:
Conformal property of the Finsler space {${\cal F}_{SR}$} and
  extension of electromagnetic field equations. \it Reports on Math.
  Phys. \bf 45 \rm (2000), 155-169.

\bibit[87]G.S. Asanov.
The Finsler-type recasting of Lorentz transformations. The
  SR Finslerian metric function and Hamiltonian function. The light
  signal velocity case. implications. In:
Proceedings of Conference \it ``Physical Interpretation of
Relativity Theory" \rm, London, Sunderland, 2000, pp. 16--40.

\bibit[88]
G.S. Asanov.
Finslerian future-past asymmetry. \it Reports on
  Math. Phys. \bf 46 \rm (2000) 100-110.

\bibit[89] G.S. Asanov: \it Finsler Geometry, Relativity and Gauge
 Theories, \rm D.~Reidel Publ. Comp., Dordrecht 1985.

\bibit[90] H. Rund: \it The Differential Geometry of Finsler spaces, \rm
Springer-Verlag, Berlin 1959.

\bibit[91] R. S. Ingarden and L. Tamassy: \it Rep. Math. Phys.
\bf32 \rm(1993), 11.

\bibit[92] R. S. Ingarden: In: \it Finsler Geometry \rm (Contemporary
Mathematics, v.~196), American Math. Soc., Providence 1996, pp.~213--223.

\bibit[93] D.~Bao, S. S. Chern, and Z. Shen (eds.): \it Finsler Geometry \rm
(Contemporary Mathematics, v.~196), American Math. Soc., Providence 1996.

\bibit[94] D.~Bao, S. S. Chern, and Z. Shen: \it An Introduction to Riemann-Finsler
Geometry. \rm N.Y., Berlin, Springer.

\bibit[95] J. Alspector et al., \it Phys. Rev. Lett. \bf36 \rm(1976), 837.

\bibit[96] G.R. Kalbfleisch et al., \it Phys. Rev. Lett. \bf43 \rm(1976), 1361.

\bibit[97] V. Ammosov and G. Volkov: ``Can neutrinos probe extra dimensions?"
hep-ph/0008032.

\end {document}